\documentclass[journal,twocolumn]{IEEEtran}
\usepackage[cmex10]{amsmath}
\usepackage{amssymb,amsthm,amsfonts,mathrsfs,xparse}
\usepackage{graphicx}
\usepackage{dsfont,enumitem,stackrel}
\usepackage{color}
\usepackage{mathtools,amssymb,bm,mathabx}
\usepackage{amstext}
\usepackage{array}
\usepackage{algorithmicx}
\usepackage[ruled]{algorithm}
\usepackage{algpseudocode}
\usepackage{algpascal}
\usepackage{algc}
\usepackage{cases}
\usepackage{pgfplots}
\usepackage{accents}
\usepackage{caption}
\usepackage{float}
\usepackage{cite}
\usepackage [autostyle, english = american]{csquotes}
\usepackage{hyperref}
\theoremstyle{remark}

\newcommand\ASTART{\bigskip\noindent\begin{minipage}[b]{0.5\linewidth}}
	
	\newcommand\AENDSKIP{\end{minipage}\bigskip}
\newcommand\AEND{\end{minipage}}
\ifCLASSOPTIONcaptionsoff
\usepackage[nomarkers]{endfloat}
\let\MYoriglatexcaption\caption
\renewcommand{\caption}[2][\relax]{\MYoriglatexcaption[#2]{#2}}
\fi
\newcommand{\RN}[1]{%
	\textup{\uppercase\expandafter{\romannumeral#1}}%
}

\def\change{black}
\theoremstyle{plain}
\newtheorem{thm}{\textbf{Theorem}}

\theoremstyle{definition}
\newtheorem{defn}{\textbf{Definition}}

\theoremstyle{remark}
\newtheorem{rem}{\bf Remark}
\newtheorem*{sketch}{\bf Proof sketch }

\newcommand*{\rom}[1]{\expandafter\@slowromancap\romannumeral #1@}

\graphicspath{{./Figures/}} 

\begin{document}
%
%\onecolumn
\title{Living near the edge: A lower-bound on the phase transition of total variation minimization}
\author{Sajad~Daei,  Farzan~Haddadi,  Arash~Amini
	\thanks{S. Daei and F. Haddadi are with the School of Electrical Engineering,
		Iran University of Science \& Technology, Tehran 16846-13114, Iran (e-mail: sajaddaeiomshi@gmail.com; farzanhaddadi@iust.ac.ir).
		A. Amini is with the Department of Electrical Engineering, Sharif University of
		Technology, Tehran 11365-8639, Iran (e-mail: arashsil@gmail.com).}
}

\maketitle

\begin{abstract}
This work is about the total variation (TV) minimization which is used for recovering gradient-sparse signals from compressed measurements. Recent studies indicate that TV minimization exhibits a phase transition behavior from failure to success as the number of measurements increases. In fact, in large dimensions, TV minimization succeeds in recovering the gradient-sparse signal with high probability when the number of measurements exceeds a certain threshold; otherwise, it fails almost certainly. Obtaining a closed-form expression that approximates this threshold is a major challenge in this field and has not been appropriately addressed yet. In this work, we derive a tight lower-bound on this threshold in case of {\color{\change} any random measurement matrix whose null space is distributed uniformly with respect to the Haar measure}. In contrast to the conventional TV phase transition results that depend on the simple gradient-sparsity level, our bound is highly affected by generalized notions of gradient-sparsity. Our proposed bound is very close to the true phase transition of TV minimization confirmed by simulation results.
\end{abstract}


% \fi
\IEEEpeerreviewmaketitle

\section{Introduction}
\IEEEPARstart{C}{ompressed} sensing (CS) has gained a lot of attention in the past decade as it provides a strategy to recover signals from undersampled measurements \cite{candes2005decoding,donoho2006most}. In mathematical terms, the measured data about a signal $\bm{x}\in\mathbb{R}^n$ is given by a collection of linear projections
\begin{align}\label{eq.lin_measure}
\bm{y}=\bm{A x}\in\mathbb{R}^m,
\end{align} 
where $\bm{A}\in\mathbb{R}^{m\times n}$ with $m\ll n$ is the measurement matrix. Without having any prior knowledge about the structure of $\bm{x}$, it is impossible to reconstruct $\bm{x}$ from $\bm{y}$. The standard prior knowledge in CS is that the signal of interest is sparse in an orthonormal basis which means that it can be expressed as the linear combination of a few basis elements. However, this simple assumption often seems irrational in practical scenarios as we deal with signals that are sparse in some overcomplete dictionary $\bm{D}\in\mathbb{R}^{n\times N}$ with $n\ll N$ which means that $\bm{x}$ can be described as $\bm{x}=\bm{D \alpha}$ for some sparse $\bm{\alpha}\in\mathbb{R}^{N}$. This setting is known as the synthesis sparsity model since it describes a way to synthesize the signal $\bm{x}$. Then, for recovering $\bm{\alpha}$ from $\bm{y}$, it is common to use the so-called $\ell_1$ synthesis problem defined as
\begin{align}\label{eq.l1syntheis}
\widehat{\alpha}=\mathop{\arg\min}_{\bm{z}\in \mathbb{R}^N}\|\bm{z}\|_1\nonumber\\
{\rm s.t.}~~\bm{y}=\bm{A D z}.
\end{align}
The reconstructed signal is then $\widehat{\bm{x}}=\bm{D}\widehat{\bm{\alpha}}$. Interestingly, the synthesis model has an analysis (and more general) counterpart which assumes that the signal of interest i.e. $\bm{x}$ is sparse after applying an operator called analysis operator $\bm{\Omega}$. The fact that $\bm{\Omega}\bm{x}$ is sparse, motivates the optimization problem 
\begin{align}\label{eq.l1analysis}
&\widehat{\bm{x}}=\mathop{\arg\min}_{\bm{z}\in\mathbb{R}^n} \|\bm{\Omega z}\|_1\nonumber\\
&{\rm s.t.}~~\bm{y}=\bm{A z},
\end{align}
which is typically referred to as $\ell_1$ analysis problem. Although the two methods $\ell_1$ synthesis and $\ell_1$ analysis perform very differently on large families of signals, the numerical results of the works \cite{candes2011compressed,elad2007analysis} show that the analysis formulation outperforms its synthesis-based counterpart in many scenarios of interest. 

An important and special case of $\ell_1$ analysis formulation \eqref{eq.l1analysis} is the case where $\bm{\Omega}$ is the finite difference matrix defined as
\begin{align}
\bm{\Omega}=\begin{bmatrix}
1& -1 & 0 & \cdots & 0 \\
0& 1 & -1 & \cdots & 0  \\
&  \ddots    & \ddots       & \ddots & \\
0 & \cdots & \cdots & 1 & -1 &
\end{bmatrix}\in\mathbb{R}^{n-1\times n}.\nonumber
\end{align}
By replacing this $\bm{\Omega}$ in the problem \eqref{eq.l1analysis}, we reach the well-known total variation (TV) minimization:
\begin{align}
\mathsf{P}_{\rm TV}:~~&\min_{\bm{z}\in \mathbb{R}^n}\|\bm{z}\|_{\mathrm{TV}}:=\|\bm{\Omega}\bm{z}\|_1=\sum_{i=1}^{n-1}|z_{i+1}-z_i|\nonumber\\
&\mathrm{s.t.}~\bm{y}_{m\times 1}=\bm{A z}.
\end{align}
TV minimization has been proven to be very effective in image processing \cite{cai2012image,rudin1992nonlinear,sidky2008image,keeling2003total,cai2015guarantees,krahmer2017total} and other fields \cite{wu2015situ,van1995total}. 
One of the important issues concerning the problem $\mathsf{P}_{\rm TV}$ is the required number of measurements i.e. the minimum $m$ that $\mathsf{P}_{\rm TV}$ needs for successful recovery. A series of works studying convex geometry approaches have found valuable results regarding the required number of measurements in recovering structured\footnote{For example, sparse signals.} signals\cite{amelunxen2013living,tropp2015convex,chandrasekaran2012convex}. Specifically, it has been shown in \cite{amelunxen2013living} that $\mathsf{P}_{\rm TV}$ undergoes a phase transition from failure to success as the number of measurements increases. This means that there exists a certain curve $\Psi(\bm{\Omega},\bm{x})$ (known as statistical dimension) in the boundary of failure and success that $\mathsf{P}_{\rm TV}$ succeeds to recover the gradient-sparse vector with probability $\frac{1}{2}$. Obtaining an expression that approximates $\Psi(\bm{\Omega},\bm{x})$ is a fundamental challenge in image processing and has not been exactly addressed yet.
In this work, we obtain a tight lower-bound on $\Psi(\bm{\Omega},\bm{x})$. As numerical results confirm, our proposed lower-bound follows the true phase transition curve very well. 
\subsection{Prior Works}
In the last few years, great works have been established for obtaining the required number of measurements in $\mathsf{P}_{\rm TV}$ (see e.g. \cite{cai2015guarantees,kabanava2015analysis,needell2013stable,poon2015role,krahmer2017total,genzel2017ell,nam2013cosparse,donoho2013accurate,daei2018sample,zhang2016precise}). 

Needell et al. in \cite{needell2013stable,needell2013near} obtain recovery guarantees for two-dimensional TV minimization. Their result is based on transforming gradient-sparse images into compressible signals in a Haar wavelet basis. Then, a modified restricted isometry property (RIP) is developed to guarantee the image recovery. Their approach does not hold for one-dimensional gradient-sparse signals. Besides, their sample complexity result is only designed for the asymptotic setting.

Nam et al. in \cite{nam2013cosparse} consider the original problem of minimizing the number of variations (NV) under some affine constraints defined as
\begin{align}\label{problem.P0}
\mathsf{P}_{\rm NV}:~~&\min_{\bm{z}\in \mathbb{R}^n}\|\bm{z}\|_{\mathrm{NV}}:=\|\bm{\Omega} \bm{z}\|_0:=\sum_{i=1}^{n-1}1_{|z_{i+1}-z_i|>0}\nonumber\\
&\mathrm{s.t.}~\bm{y}_{m\times 1}=\bm{A z},
\end{align}
where $1_{\mathcal{E}}$ denotes the indicator function of a set $\mathcal{E}$. Unfortunately, $\mathsf{P}_{\rm NV}$ is known to be NP-complete \cite[Section 4.1]{nam2013cosparse} and $\mathsf{P}_{\rm TV}$ is the closest tractable problem to $\mathsf{P}_{\rm NV}$. In fact, the TV norm sums the amplitudes of variations and in some sense, is to the NV function what the $\ell_1$ norm is to the $\ell_0$ function in the area of sparse recovery. For a $s$-gradient-sparse vector $\bm{x}\in\mathbb{R}^n$, it has been shown in \cite{nam2013cosparse} that
\begin{align}
m>2\kappa_{\bm{\Omega}}(s)
% :=2\max_{|\mathcal{S}|<n-1-s}{\rm dim}({ \rm null}(\bm{\Omega}_{\overline{\mathcal{S}}}))
\end{align}
measurements suffice for exact recovery via $\mathsf{P}_{\rm NV}$. Here, $\kappa_{\bm{\Omega}}(s)$ is a special function describing the signal manifold dimension and is obtained via a combinatorial search over all $s$-gradient-sparse signals. This result matches with the well-known fact in conventional CS (in particular sparse recovery and matrix completion) where the number of needed measurements is proportional to the signal's manifold dimension. However, the authors of \cite{giryes2015effective} have numerically shown that when we are dealing with the recovery of gradient-sparse signals via $\mathsf{P}_{\rm TV}$, the required number of measurements is not explained by the manifold dimension anymore. Somewhat interestingly, our proposed bound in Section \ref{section.mainresult} is consistent with this fact. 

Donoho et al. in \cite{donoho2013accurate} obtain the asymptotic minimax mean square error (MSE) of a TV denoiser and using numerical simulations, conjecture that it matches with the phase transition curve of $\mathsf{P}_{\rm TV}$. However, this conjecture is not proved. Moreover, the result holds only in the asymptotic case. 
%unlike $\mathsf{P}_{\rm NV}$, the required sample complexity for tractable problem $\mathsf{P}_{\rm TV}$ seems to be not well explained by the function $\kappa_{\bm{\Omega}}(s)$ or simple gradient sparsity level $s$

Cai et al. in \cite{cai2015guarantees} show that $\mathcal{O}(\sqrt{sn}\log(n))$ and $\mathcal{O}(\sqrt{sn})$ measurements are respectively sufficient and necessary for exact recovery of a given $s$-gradient-sparse vector via $\mathsf{P}_{\rm TV}$. The order of their bounds seems to be optimal. However, their (necessary and sufficient) bounds do not provide a good prediction for sample complexity in the non-asymptotic case.

In \cite[Theorem 4]{kabanava2015robust}, an upper-bound is derived for the statistical dimension (in case of TV minimization) using the results of \cite{foygel2014corrupted}. Their bound depends on the gradient-sparsity level and provides an inaccurate prediction for the required sample complexity in low-sparsity regimes.

Recently, the authors of \cite{daei2018sample} obtain a non-asymptotic sample complexity bound for $\mathsf{P}_{\rm TV}$. In contrast to the previous bounds that would depend on the gradient-sparsity $s$, their non-asymptotic bound is highly affected by the number of consecutive variations (adjacent pairs in the gradient support). Their proof approach is based on a refined analysis of \cite{foygel2014corrupted}. While their bound outperforms the previous bounds, it is still far from the statistical dimension in low-sparsity regimes.

By using a more general analysis of \cite{foygel2014corrupted}, the authors in \cite{genzel2017ell} obtain an explicit formula describing the required number of measurements. Their bound depends on the coherence structure of $\bm{\Omega}$. Again, the proposed bound does not provide an accurate prediction of the true sample complexity in low-sparsity regimes.
\subsection{Contribution}
 In this work, we obtain a lower-bound on the statistical dimension that provides the necessary number of measurements that $\mathsf{P}_{\rm TV}$ needs for exact recovery. Our bound is very close to the true sample complexity even in low-sparsity levels and numerical experiments in Section\ref{section.simulation} show that it is tight in the asymptotic case. Our bound depends on the number of the consecutive, individual and tail-end variations of the interested signal. We hope that our bound sheds more light on the effective parameters in the statistical dimension in case of TV minimization. It is worth mentioning that the only lower-bound on the statistical dimension in the literature is established in \cite[Theorem 2.1 Part b]{cai2015guarantees}, and therefore served as an object of comparison in Sections \ref{section.mainresult} and \ref{section.simulation}.   
\subsection{Outline of the paper}
The paper is organized as follows. First, we review a few basic concepts in convex geometry. Section \ref{section.mainresult} amounts to our main result. Section \ref{section.simulation} is about numerical experiments that verify our theoretical findings. Finally, the paper is concluded in Section \ref{section.conclusion}. 
\subsection{Notation}
Throughout the paper, scalars are denoted by lowercase letters, vectors by lowercase boldface letters, and matrices by uppercase boldface letters. The $i$th element of a vector $\bm{x}$ is given either by ${x}(i)$ or $x_i$. $(\cdot)^\dagger$ denotes pseudo inverse. We reserve calligraphic uppercase letters for sets (e.g. $\mathcal{S}$). The cardinality of a set $\mathcal{S}$ is denoted by $|\mathcal{S}|$. {\color{\change}The complement of a set $\mathcal{S}$ in $\{1,..., n\}$ is denoted by ${\overline{\mathcal{S}}}$}. $\mathcal{C}^\circ$ denotes the polar of a cone $\mathcal{C}$. The symbol ${\rm cone}(\cdot)$ signifies the conic hull of a set. Null space of a matrix is shown by $\mathrm{null}(\cdot)$. For a matrix $\bm{A}$, $\bm{A}_{\mathcal{S}}$ means the matrix $\bm{A}$ restricted to the rows indexed by $\mathcal{S}$. We denote i.i.d. standard Gaussian random vector by $\bm{g}$. Lastly, $\lor$ means logical \lq\lq or \rq\rq.
\section{Convex Geometry}\label{section.convexgeom} 
%\subsection{Subdifferential}
%Subdifferential is a generalized concept of gradient for non-smooth functions. The subdifferential of homogeneous (i.e. $f(\alpha \bm{z})=|\alpha|f(\bm{z})$) and sub-additive (i.e. $f(\bm{x}+\bm{y})\le f(\bm{x})+f(\bm{y})$)
%The descent cone of a proper convex function $f:\mathbb{R}^n\rightarrow \mathbb{R}\cup \{\pm\infty\}$ at point $\bm{x}\in \mathbb{R}^n$ is the set of directions from $\bm{x}$ that do not increase $f$:
%\begin{align}\label{eq.descent cone}
%\mathcal{D}(f,\bm{x})=\bigcup_{t\ge0}\{\bm{z}\in\mathbb{R}^n: f(\bm{x}+t\bm{z})\le f(\bm{x})\}\cdot
%\end{align}
%The descent cone of a convex function is a convex set. There is a famous duality \cite[Ch. 23]{rockafellar2015convex} between decent cone and subdifferential of a convex function  given by:
%\begin{align}\label{eq.D(f,x)}
%\mathcal{D}^{\circ}(f,\bm{x})=\mathrm{cone}(\partial f(\bm{x})):=\bigcup_{t\ge0}t.\partial f(\bm{x}).
%\end{align}
\subsection{Statistical Dimension}
\begin{defn}{Statistical Dimension}\cite{amelunxen2013living}:
	Statistical dimension is a generalized concept of subspace dimension in the class of convex cones. Intuitively, it measures the size of a cone. More precisely, for a closed convex cone $\mathcal{C}\subseteq\mathbb{R}^n$, statistical dimension of $\mathcal{C}$ is defined as:
	\begin{align}\label{eq.statisticaldimension}
	\delta(\mathcal{C}):=\mathds{E}\inf_{\bm{z}\in\mathcal{C}^{\circ}}\|\bm{g}-\bm{z}\|_2^2,
	\end{align}
	where $\bm{g}$ is a random vector in $\mathbb{R}^n$ chosen from Gaussian ensemble with independent entries.
%	where, $\mathcal{P}_\mathcal{C}(\bm{x})$ is the projection of $\bm{x}\in \mathbb{R}^n$ onto the set $\mathcal{C}$ defined as: $\mathcal{P}_\mathcal{C}(\bm{x})=\underset{\bm{z} \in \mathcal{C}}{\arg\min}\|\bm{z}-\bm{x}\|_2$.
\end{defn}
%Since calculating the statistical dimension is difficult, it is common to use the upper bound
%\begin{align}
%\delta(\mathcal{D}(,\bm{x}))
%\end{align}
\subsection{Linear Inverse Problems and Sample Complexity}
In \cite{amelunxen2013living}, it is proved that any random convex optimization problem $\mathsf{P}_f$:
\begin{align}\label{eq.mainprob}
\mathsf{P}_f:~~~\min_{\bm{x}\in \mathbb{R}^n} ~f(\bm{x})~~~~
\mathrm{s.t.} ~~\bm{Ax}=\bm{b}.
\end{align}
undergoes a phase transition between success and failure as the number of measurements increases. The location of this transition (the boundary of success and failure) is specified by the statistical dimension of the descent cone of $f$ at $\bm{x}\in \mathbb{R}^n$ i.e. $\delta(\mathcal{D}(f,\bm{x}))$\cite{amelunxen2013living}. Here, $\mathcal{D}(f,\bm{x})$ is defined as the set of directions toward which $f$ is decreased and is given by
\begin{align}\label{eq.descent cone}
\mathcal{D}(f,\bm{x})=\bigcup_{t\ge0}\{\bm{z}\in\mathbb{R}^n: f(\bm{x}+t\bm{z})\le f(\bm{x})\}.
\end{align}
There is also a well-known fact between the descent cone and the subdifferential ( c.f.\cite[Chapter 23]{rockafellar2015convex}) given by
\begin{align}
\mathcal{D}(f,\bm{x})^\circ={\rm cone}(\partial f (\bm{x})).
\end{align}
By using this fact and \eqref{eq.statisticaldimension}, one may write:
\begin{align}
&\delta(\mathcal{D}(f,\bm{x}))=\mathds{E}\inf_{\bm{z}\in{\rm cone}(\partial f(\bm{x}))}\|\bm{g}-\bm{z}\|_2^2=\nonumber\\
&\mathds{E}\inf_{t\ge0}\inf_{\bm{z}\in\partial f(\bm{x})}\|\bm{g}-t\bm{z}\|_2^2.
\end{align}
Calculating statistical dimension has been a difficult task in the literature and therefore, it is commonly approximated with the expression
\begin{align}\label{eq.Bu}
B_u:=\inf_{t\ge0}\mathds{E}\inf_{\bm{z}\in\partial f(\bm{x})}\|\bm{g}-t\bm{z}\|_2^2,
\end{align}
which is first proposed by Stojnic \cite{stojnic2009various} in the context of $\ell_1$ minimization.
In \cite[Theorem 2]{daei2019error}, it has been shown that $B_u$ approximates the statistical dimension well for a large class of structure inducing functions in particular $f=\|\cdot\|_{\rm TV}$.
\section{Main result}\label{section.mainresult}
\begin{figure}[t]
	\hspace*{-.9cm}
	\centering
	\includegraphics[scale=.23]{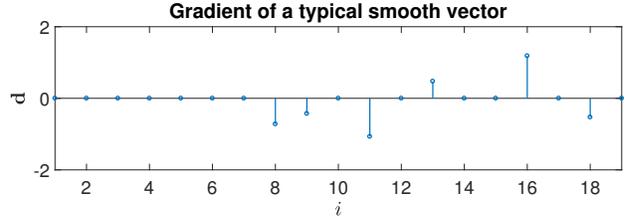}
	\caption{This plot shows the discrete gradient of a typical gradient-sparse vector $\bm{x}\in\mathbb{R}^{20}$ with parameters $s_1^+=1$, $s_1^{-}=0$, $s_2=10$, and $s_3=0$.}
	\label{fig.typical}
\end{figure}
Before stating our main result, we need to define some parameters regarding gradient-sparse signals which are required in our analysis.
\begin{defn}(Consecutive, individual and tail-end variations)\label{def.var}
Consider a gradient-sparse signal $\bm{x}\in\mathbb{R}^n$ with gradient $\bm{d}:=\bm{\Omega}\bm{x}$ and gradient support $\mathcal{S}$. The consecutive variations correspond to the adjacent pairs in $\mathcal{S}$ defined as
\begin{align}
&\mathcal{S}_1:=\{i\in[n-1]:i\in\mathcal{S},~i-1\in\mathcal{S}\}.
\end{align}
The set $\mathcal{S}_1$ can be divided into two sets:
\begin{align}
&\mathcal{S}_1^{+}:=\nonumber\\
&\{i\in[n-1]:i\in\mathcal{S},~i-1\in\mathcal{S},{\rm sgn}(d_i){\rm sgn}(d_{i-1})>0\},\nonumber\\
&\mathcal{S}_1^{-}:=\nonumber\\
&\{i\in[n-1]:i\in\mathcal{S},~i-1\in\mathcal{S},{\rm sgn}(d_i){\rm sgn}(d_{i-1})<0\},
\end{align}
which are interpreted as the consecutive variations (adjacent pairs in $\mathcal{S}$) with the same and opposite signs, respectively. We define the individual variations by the sets
\begin{align}
&\mathcal{S}_2:=\{i\in[n-1]:i\in\mathcal{S},~i-1\in\overline{\mathcal{S}}\},\nonumber\\
&\mathcal{S}_2^\prime:=\{i\in[n-1]: i\in\overline{\mathcal{S}},~i-1\in\mathcal{S}\}.
\end{align}
The variations in the left and right ends of the signal are called tail-end variations which are defined by the set
\begin{align}
&\mathcal{S}_3:=\nonumber\\
&\{i\in[n-1]:i\in\mathcal{S},~i-1\notin[n-1]\lor i\in\mathcal{S},~i+1\notin[n-1]\}.
\end{align}
We also show the number of consecutive, individual and tail-end variations respectively by
\begin{align}
&s_1:=|\mathcal{S}_1|=|\mathcal{S}_1^{+}|+|\mathcal{S}_1^{-}|:=s_1^{+}+s_1^{-}\nonumber\\
&s_2:=|\mathcal{S}_2|+|\mathcal{S}_2^{\prime}|,\nonumber\\
&s_3:=|\mathcal{S}_3|\nonumber.
\end{align}
\end{defn}
As an illustrative example, in Figure \ref{fig.typical}, the discrete gradient $\bm{d}:=\bm{\Omega x}$ of a typical gradient-sparse signal $\bm{x}\in\mathbb{R}^{20}$ is depicted. There is one pair of elements in $\bm{d}$ both of which belong to the gradient support (alternatively representing consecutive variations of $\bm{x}$) and have the same sign. Thus, $s_1^{+}=1$ and $s_2^{-}=0$. Also, there are $10$ elements for which $i\in\mathcal{S}, i-1\notin\overline{\mathcal{S}}$ or $i\in\overline{\mathcal{S}}, i-1\notin\mathcal{S}$ (alternatively representing individual variations in $\bm{x}$). As a result, $s_2=10$. Lastly, there are no elements in the tail-ends of $\bm{d}$. This means that the first and last elements of $\bm{x}$ include no variations and $s_3=0$. 

In the following theorem, we propose a lower-bound on $\delta(\mathcal{D}(\|\cdot\|_{\rm TV},\bm{x}))$. This lower-bound depends on the associated parameters in Definition \ref{def.var}
\begin{thm}\label{thm.main}
Let $\bm{x}\in\mathbb{R}^n$ be a gradient-sparse signal with gradient $\bm{d}:=\bm{\Omega x}$, and gradient support $\mathcal{S}$. Assume that $\bm{x}$ has $s_1=s_1^{+}+s_1^{-}$, $s_2$, and $s_3$ consecutive, individual and tail-end variations, respectively as defined in Definition \ref{def.var}. Let  $\bm{A}\in\mathbb{R}^{m\times n}$ be {\color{\change} a random matrix  whose null space is uniformly distributed with respect to the Haar measure}. Consider $\bm{y}_{m\times 1}=\bm{A x}$ as the vector of measurements. Then,
\begin{align}
\delta(\mathcal{D}(\|\cdot\|_{\rm TV},\bm{x}))\ge \inf_{t\ge0}\Psi_t(s_1^{+},s_1^{-},s_2,s_3):=\widehat{m}_{\rm TV},
\end{align}
where
\begin{align}
&\Psi_t(s_1^{+},s_1^{-},s_2,s_3)=s_1^{+}+s_1^{-}(1+4t^2)\nonumber\\
&+s_2\phi_1(t,t)+(n-2-s_1^{+}-s_1^{-}-s_2)\phi_2(2t)+\nonumber\\
&s_3(1+t^2)+(2-s_3)\phi_2(t),
\end{align}
and
\begin{align}\label{eq.gradientsupport}
%&s_1^{+}=\Big|\{i\in[n-1]:i\in\mathcal{S},~i-1\in\mathcal{S},{\rm sgn}(d_i){\rm sgn}(d_{i-1})>0\}\Big|,\nonumber\\
%&s_1^{-}=\nonumber\\
%&\Big|\{i\in[n-1]:i\in\mathcal{S},~i-1\in\mathcal{S},{\rm sgn}(d_i){\rm sgn}(d_{i-1})<0\}\Big|,\nonumber\\
%&s_2=\Big|\{i\in[n-1]:i\in\mathcal{S},~i-1\in\overline{\mathcal{S}}\lor i\in\overline{\mathcal{S}},~i-1\in\mathcal{S}\}\Big|,\nonumber\\
%&s_3=\Big|\{i\in[n-1]:i\in\mathcal{S},~i-1\notin[n-1]\lor \nonumber\\
%&~~~~~~~~~~~~~~~~~~~~~~~~~ i\in\mathcal{S},~i+1\notin[n-1] \}\Big|,\nonumber\\
&\phi_1(a,b):=\frac{1}{\sqrt{2\pi}}\int_{b}^{\infty}(u-b)^2[e^{-\frac{(u-a)^2}{2}}+e^{-\frac{(u+a)^2}{2}}]du,\nonumber\\
&\phi_2(z):=\sqrt{\frac{2}{\pi}}\int_{z}^{\infty}(u-z)^2e^{-\frac{u^2}{2}}du,\nonumber\\
%&\sigma_1=\frac{|\mathcal{S}_1|}{n-1},~\sigma_2=\frac{|\mathcal{S}_2|}{n-1}, ~\sigma_3=\frac{|\mathcal{S}_3\cup\mathcal{S}_4|}{n-1},\sigma_4=\frac{|\mathcal{S}_6\cup\mathcal{S}_7|}{n-1},\nonumber\\
%&\overline{\sigma}_4=\frac{|\mathcal{\overline{S}}_6\cup\mathcal{\overline{S}}_7|}{n-1}.\nonumber\\
%&s=|\mathcal{S}_1\cup\mathcal{S}_2\cup\mathcal{S}_3\cup\mathcal{S}_6|.
\end{align}
and thus if $m\le \widehat{m}_{\rm TV}$, then with probability at least $1-4{\rm e}^{-\tfrac{(\widehat{m}_{\rm TV}-m)^2}{16\widehat{m}_{\rm TV}}}$, $\mathsf{P}_{\rm TV}$ fails to recover $\bm{x}$.
%In (\ref{eq.gradientsupport}), $\mathcal{S}_1\cup\mathcal{S}_2$ and $\mathcal{S}_3\cup\mathcal{S}_4$ refer to consecutive and individual support, respectively.	
\end{thm}
\begin{sketch}
As discussed in \eqref{eq.Bu}, we intend to find a lower-bound for
\begin{align}
\inf_{t\ge0}\mathds{E}\inf_{\bm{z}\in\partial\|\cdot\|_{\mathrm{TV}}(\bm{x})}\|\bm{g}-t\bm{z}\|_2^2.
\end{align}
The expression inside the latter infimum is formed of a few summands. Each summand separately has a unique minimizer over the set $\partial \|\cdot\|_{\rm TV}(\bm{x})$. By passing this infimum through each summand (this leads to a lower-bound) and then applying expectation, we reach a closed-form expression for each summand and thereby a strictly convex function with respect to $t$ for the expression inside the former infimum. 
\end{sketch}

\textit{Discussion}. One of the key properties of Theorem \ref{thm.main} is that, unlike the most works in TV minimization, the widely-used concept of gradient sparsity $s:=|\mathcal{S}|$ does not directly explain our proposed bound. In fact, it seems that the statistical dimension in case of TV minimization i.e. $\delta(\mathcal{D}(\|\cdot\|_{\rm TV},\bm{x}))$, depends on generalized concepts of gradient-sparsity: namely the number of consecutive, individual, and tail-end variations. 
{\color{\change} 
To examine how these parameters affect our bound $\widehat{m}_{\rm TV}$, we designed a numerical experiment in Table \ref{table.interpretation} and considered different values for $s_1^{+}$, $s_1^{-}$, $s_2$ and $s_3$. We observe from Table \ref{table.interpretation} that the number of consecutive variations with negative signs, i.e. $s_1^{-}$, has the highest impact on $\widehat{m}_{\rm TV}$. In other words, for a fixed number of variations ($s=\|\bm{\Omega} \bm{x}\|_0$), recovering a highly oscillating signal is harder (alternatively needs more measurements) than the one with well-separated variations.
}
\begin{rem}(Prior work)
In \cite[Theorem 2.1]{cai2015guarantees}, it has been proved that 
\begin{align}
&\frac{9\sqrt{s n}}{50\pi}-\frac{12}{5\pi}\le\delta(\mathcal{D}(\|\cdot\|_{\rm TV},\bm{x}))\le\nonumber\\
&\sqrt{32}(2\sqrt{5}+\sqrt{10})^2\sqrt{ns}\log(2n)+1.
\end{align}
Their proof approach is based on estimating the statistical dimension of a certain set using a wavelet-based argument. Since the lower and upper-bound are in the same order (up to a log factor), their approach is optimal in the asymptotic case ($n\rightarrow \infty$). However, in the non-asymptotic case, it demonstrates poor prediction of the true sample complexity.
\end{rem}
\begin{figure}[t]
	\hspace*{-0.0cm}
	\centering
	\includegraphics[scale=.27]{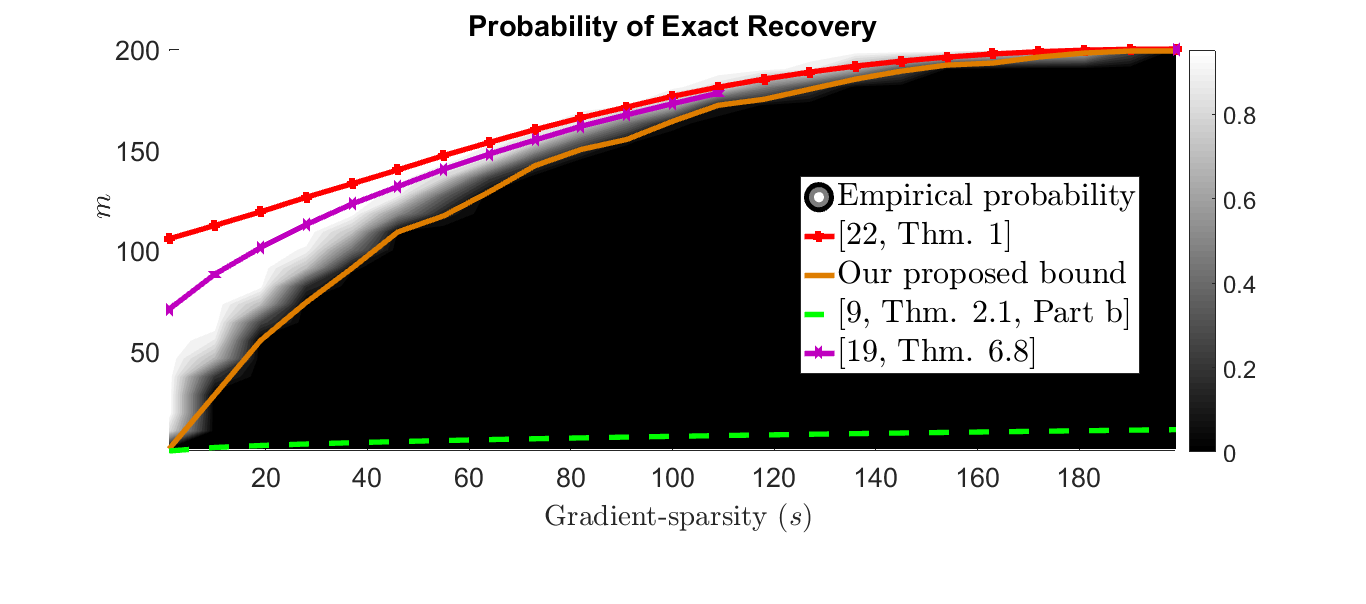}
	\caption{This figure shows the phase transition of $\mathsf{P}_{\rm TV}$ in case of $n=200$. The {\color{\change}orange} curve is our proposed bound. The bounds in \cite[Theorem 1]{daei2018sample} and \cite[Theorem 2.1 part b]{cai2015guarantees} are depicted by dashed and dotted line, respectively. The {\color{\change}orange}, purple, and red curves are obtained by computing the empirical mean of the sample complexity for each $s$. The brightness of figure in each pair $(s,m)$, shows the empirical probability of success (black=$0\%$, white=$100\%$).}
	\label{fig.phase1}
\end{figure}

\section{Simulations}\label{section.simulation}
\begin{figure}[t]
	\hspace*{-0.3cm}
	\centering
	\includegraphics[scale=.26]{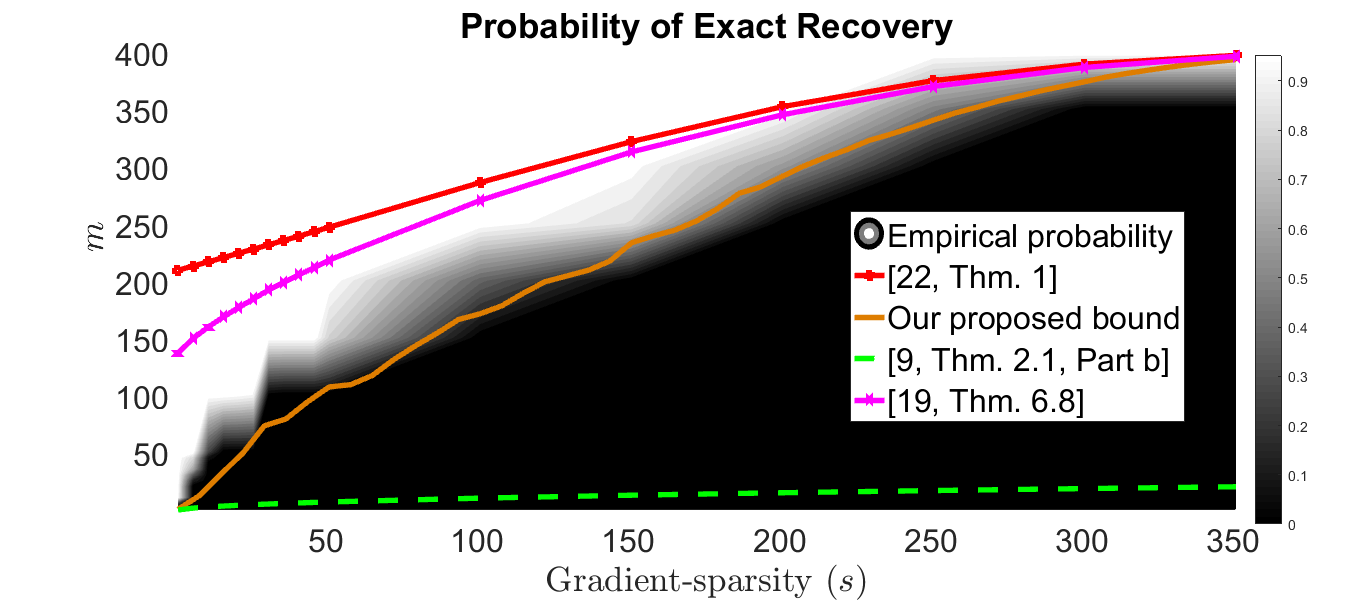}
	\caption{This figure shows the phase transition of $\mathsf{P}_{\rm TV}$ in case of $n=400$. The {\color{\change}orange} curve is our proposed bound. The bounds in \cite[Theorem 1]{daei2018sample} and \cite[Theorem 2.1 part b]{cai2015guarantees} are depicted by dashed and dotted line, respectively. The {\color{\change}orange}, purple, and red curves are obtained by computing the empirical mean of the sample complexity for each $s$. The brightness of figure in each pair $(s,m)$, shows the empirical probability of success (black=$0\%$, white=$100\%$).}
	\label{fig.phase2}
\end{figure}

In this section, we evaluate how our proposed bound in Theorem \ref{thm.main} predicts the phase transition of $\mathsf{P}_{\rm TV}$. For each $m$ and $s$, we repeat the following steps $50$ times:
\begin{itemize}
	\item Select a random subset $\mathcal{S} \subseteq \{1,..., n-1\}$ with $|\mathcal{S}|=s$. 
	\item Generate a vector $\bm{x}\in {\rm null}(\bm{\Omega}_{\overline{\mathcal{S}}})$ whose gradient is supported on $\mathcal{S}$.
	\item Construct the vector $\bm{y}=\bm{A x}$ where $\bm{A}$ is an i.i.d. Gaussian matrix {\color{\change}(the null space of Gaussian matrices is distributed uniformly with respect to the Haar measure)}.
	\item Solve $\mathsf{P}_{\rm TV}$ to obtain an estimate $\widehat{\bm{x}}$.
	\item Declare success if $\|\bm{x}-\widehat{\bm{x}}\|_2\le 10^{-6}$.
\end{itemize} 
Figures \ref{fig.phase1}, and \ref{fig.phase2} show the empirical probability of successful recovery obtained from $50$ Monte Carlo iterations in case of $n=200$ and $n=400$, respectively. Notice that since the sample complexity bounds in \cite[Theorem 1]{daei2018sample}, \cite[Theorem 1]{genzel2017ell} and $\widehat{m}_{\rm TV}$ do not directly depend on the gradient-sparsity $s$, we depict the empirical mean of the bounds over $300$ iterations, for each $s$. As it turns out from Figures \ref{fig.phase1}, and \ref{fig.phase2} our proposed bound in Theorem \ref{thm.main} predicts the statistical dimension well; the lower-bound \cite[Theorem 2.1]{cai2015guarantees} does not seem to be exact; and the upper-bounds \cite[Theorem 1]{daei2018sample} and \cite[Theorem 6.8]{genzel2017ell} fail to explain the true sample complexity in low-sparsity levels.

\begin{table}[t]
	\centering
	{\color{\change}
		\begin{tabular}{ |c|c|c|c|c|c|c|}%{p{0.25\linewidth}p{0.25\linewidth}p{0.25\linewidth}}
			\hline
			$n$&$s$&$s_1^{+}$&$s_1^{-}$&$s_2$ & $s_3$&$\widehat{m}_{\rm TV}$\\ 
			\hline
			$100$&$10$&$9$ & $0$ & $2$&$0$&$10$ \\
			\hline
			$100$&$10$&$0$ & $9$ & $2$&$0$&$32.04$\\ 
			\hline
			$100$&$10$&$0$ & $8$ & $2$&$2$&$31.74$ \\
			\hline
			$100$&$10$&$0$ & $9$ & $1$&$1$&$32.584$\\
			\hline
			$100$&$10$&$9$ & $0$ & $1$&$1$&$12.33$ \\
			\hline
			$100$&$10$&$0$ & $0$ & $20$&$0$&$10$\\
			\hline
			$100$&$10$&$0$ & $1$ & $17$&$1$&$16.53$\\
			\hline
			$100$&$10$&$4$ & $5$ & $2$&$0$&$25.54$\\
			\hline
		\end{tabular}
	}
	\caption{{\color{\change}In this table, we examine the impact of $s_1^{+}$, $s_1^{-}$, $s_2$ and $s_3$ on our bound $\widehat{m}_{\rm TV}$ in Theorem \ref{thm.main}. We observe that $s_1^{-}$ has the highest impact.}}\label{table.interpretation}
\end{table}

\section{Conclusion}\label{section.conclusion}
In this work, we obtained a lower-bound on the statistical dimension in case of TV minimization. This lower-bound provides the necessary number of measurements that $\mathsf{P}_{\rm TV}$ needs for successful recovery. Our bound depends on the number of consecutive, individual and tail-end variations of the signal and precisely captures the location of the TV phase transition. In fact, it seems that these quantities specify the effective parameters of the statistical dimension for gradient-sparse signals.
%\vspace{-3cm}
\appendices
\section{Proof of Theorem \ref{thm.main}}\label{proof.thm1}
%\vspace{-1.5cm}
\begin{proof}
Recall that the sets $\mathcal{S}_1^{+}$, $\mathcal{S}_1^{-}$, $\mathcal{S}_2$, $\mathcal{S}_2^\prime$, $\mathcal{S}_3$ are defined in Definition \ref{def.var}. Moreover, define the set of adjacent pairs in $\overline{\mathcal{S}}$ as
\begin{align}
%&\mathcal{S}_1^{+}:=\nonumber\\
%&\{i\in[n-1]:i\in\mathcal{S},~i-1\in\mathcal{S},{\rm sgn}(d_i){\rm sgn}(d_{i-1})>0\},\nonumber\\
%&\mathcal{S}_1^{-}:=\nonumber\\
%&\{i\in[n-1]:i\in\mathcal{S},~i-1\in\mathcal{S},{\rm sgn}(d_i){\rm sgn}(d_{i-1})<0\},\nonumber\\
%&\mathcal{S}_2:=\{i\in[n-1]:i\in\mathcal{S},~i-1\in\overline{\mathcal{S}}\},\nonumber\\
%&\mathcal{S}_2^\prime:=\{i\in[n-1]: i\in\overline{\mathcal{S}},~i-1\in\mathcal{S}\},\nonumber\\
%&\mathcal{S}_3:=\nonumber\\
%&\{i\in[n-1]:i\in\mathcal{S},~i-1\notin[n-1]\lor i\in\mathcal{S},~i+1\notin[n-1]\},\nonumber\\
%&\mathcal{S}_3^{\prime}:=\nonumber\\
%&\{i\in[n-1]:i\in\overline{\mathcal{S}},~i-1\notin[n-1]\lor i\in\overline{\mathcal{S}},~i+1\notin[n-1]\},\nonumber\\
&\mathcal{S}_4:=\{i\in[n-1]:~i\in\overline{\mathcal{S}},~i-1\in\overline{\mathcal{S}}\},
\end{align}
which are used in the proof.

Since statistical dimension is approximately equal to $B_u$ in \eqref{eq.Bu}, we find a lower-bound for
%\vspace{-1cm}
\begin{align}\label{eq.mTVweighted}
&B_u:= \inf_{t\ge0}\mathds{E}\mathrm{dist}^2(\bm{g},t\partial\|\cdot\|_{\mathrm{TV}}( \bm{x})).
\end{align}
The first step is to calculate $\partial\|\cdot\|_{\mathrm{TV}}(\bm{x})$. From the chain rule lemma of subdifferential \cite[Chapter 23]{rockafellar2015convex}, we have:
\begin{align}\label{eq.subdiffTVW}
&\partial\|\cdot\|_{\mathrm{TV}}(\bm{x})=\bm{\Omega}^T\partial\|\cdot\|_{1}(\bm{d})=\nonumber\\
&\bm{\Omega}^T\left\{\bm{z}\in\mathbb{R}^{n-1}:\begin{array}{lr}
z_i={\rm sgn}(d_i), &  i\in \mathcal{S}\\
|z_i|\le 1, & \mathrm{o.w.}
\end{array}\right\}.
\end{align}
To calculate \eqref{eq.mTVweighted}, regarding (\ref{eq.subdiffTVW}), we compute the distance of the dilated subdifferential of the descent cone of TV norm at $\bm{x}\in\mathbb{R}^n$ from a standard Gaussian vector $\bm{g}\in\mathbb{R}^n$ which is given by:
\begin{align}\label{eq.dist1}
	&\mathrm{dist}^2(\bm{g},t\partial\|\cdot\|_{\mathrm{TV}}(\bm{x}))=\inf_{\bm{z}\in\partial\|\cdot\|_{\mathrm{TV}}(\bm{x})}\|\bm{g}-t\bm{z}\|_2^2=\nonumber\\
	&\inf_{\bm{z}\in\partial\|\cdot\|_{\mathrm{TV}}(\bm{x})}\sum_{i=1}^{n}(g_i-t(\Omega^Tz)_i)^2=\nonumber\\
	&\inf_{\|\bm{z}\|_{\infty}\le 1}\sum_{i=1}^{n}\big(g_i-t\sum_{j\in\mathcal{S}}\Omega(j,i){\rm sgn}(d_j)
	\nonumber\\
	&-t\sum_{j\in\overline{\mathcal{S}}}\Omega(j,i)z(j)\big)^2.
\end{align}
Each row of the finite difference operator $\bm{\Omega}$ includes a pair of $\{+1,-1\}$ and is zero elsewhere, i.e.,
\begin{align}
 \Omega(j,i) = \left\{\begin{array}{lr}
1, &  j=i\\
-1, & j=i-1
\end{array}\right\}
\end{align}
By using the latter property, the relation \eqref{eq.dist1} reads
\begin{align}\label{eq.dist2}
&\inf_{\|\bm{z}\|_{\infty}\le 1}\sum_{i=1}^{n}\bigg(g_i-
	t{\rm sgn}(d_i)1_{i\in\mathcal{S}}+t{\rm sgn}(d_{i-1})1_{i-1\in\mathcal{S}}\nonumber\\
	&-tz(i)1_{i\in\overline{\mathcal{S}}}+tz(i-1)1_{i-1\in\overline{\mathcal{S}}}\bigg)^2.
\end{align}
By passing the infimum through the summation, we have a lower-bound on \eqref{eq.dist2} as follows:
	\begin{align}\label{eq.dist3}
	&\mathrm{dist}^2(\bm{g},t\partial\|\cdot\|_{\mathrm{TV}}(\bm{x}))\ge \nonumber\\	
	&\sum_{i\in\mathcal{S}_1^{+}\cup\mathcal{S}_1^{-}}(g_i-t{\rm sgn}(d_i)+t{\rm sgn}(d_{i-1}))^2\nonumber\\
	&+\sum_{i\in\mathcal{S}_2}\inf_{\|\bm{z}\|_{\infty}\le 1}(g_i-t{\rm sgn}(d_i)+tz(i-1))^2+\nonumber\\
	&\sum_{i\in\mathcal{S}_2^\prime}\inf_{\|\bm{z}\|_{\infty}\le 1}(g_i+t{\rm sgn}(d_{i-1})-tz(i))^2+\sum_{i\in\mathcal{S}_4}\inf_{\|\bm{z}\|_{\infty}\le 1}(g_i-tz(i)\nonumber\\
	&+tz(i-1))^2+(g_1-t{\rm sgn}(d_1))^2 1_{1\in\mathcal{S}}+\nonumber\\
	&\inf_{\|\bm{z}\|_{\infty}\le 1}(g_1-tz(1))^2 1_{1\in\overline{\mathcal{S}}}+(g_n-t{\rm sgn}(d_{n-1}))^2 1_{n-1\in\mathcal{S}}+\nonumber\\
	&\inf_{\|\bm{z}\|_{\infty}\le 1}(g_n-tz(n-1))^21_{n-1\in\overline{\mathcal{S}}}.
	\end{align}
	Now, we investigate the minimizations in \eqref{eq.dist3}, one by one. First, it holds that
	\begin{align}
	&\inf_{\|\bm{z}\|_{\infty}\le 1}(g_i-t{\rm sgn}(d_i)+tz(i-1))^2=\nonumber\\
	&\inf_{|z(i-1)|\le 1}(g_i-t{\rm sgn}(d_i)+tz(i-1))^2\nonumber\\
	&\stackrel{(\RN{1})}{=}(|g_i-t{\rm sgn}(d_i)|-t)_{+}^2,
	\end{align}
	where the equality $(\RN{1})$ is since the optimal objective function of the problem
	$$\inf_{|z(i-1)|\le 1}(g_i-t{\rm sgn}(d_i)+tz(i-1))^2,$$
	occurs either by the boundaries imposed by the feasible set $|z(i-1)|\le 1$ or equals zero. With a similar reason, we have
	\begin{align}
	\inf_{\|\bm{z}\|_{\infty}\le 1}(g_i+t{\rm sgn}(d_{i-1})-tz(i))^2=(|g_i+t{\rm sgn}(d_{i-1})|-t)_{+}^2,
	\end{align}
	and
	\begin{align}
	&\inf_{\|\bm{z}\|_{\infty}\le 1}(g_i-tz(i)+tz(i-1))^2=\nonumber\\
	&\inf_{|z(i)|\le 1, |z(i-1)|\le 1}(g_i-tz(i)+tz(i-1))^2= (|g_i|-2t)_+^2.
	\end{align}
	Introduce the above expressions into \eqref{eq.dist1} to reach
	\begin{align}
	&\mathrm{dist}^2(\bm{g},t\partial\|\cdot\|_{\mathrm{TV}}(\bm{x}))\ge \nonumber\\
	&=\sum_{i\in\mathcal{S}_1^{+}\cup\mathcal{S}_1^{-}}(g_i-t{\rm sgn}(d_i)+t{\rm sgn}(d_{i-1}))^2\nonumber\\
	&+\sum_{i\in\mathcal{S}_2}(\zeta_1-t)_{+}^2+\sum_{i\in\mathcal{S}_2^\prime}(\zeta_2-t)_{+}^2\nonumber\\
	&+\sum_{i\in\mathcal{S}_4}(|g_i|-t-t)_{+}^2+(g_1-t{\rm sgn}(d_1))^21_{1\in\mathcal{S}}+\nonumber\\
	&(|g_1|-t)_{+}^2 1_{1\in\overline{\mathcal{S}}}+(g_n-t{\rm sgn}(d_{n-1}))^2 1_{n-1\in\mathcal{S}}+\nonumber\\
	&(|g_n|-t)^21_{n-1\in\overline{\mathcal{S}}},\nonumber\\
	\end{align}
	where $\zeta_1=|g_i-t{\rm sgn}(d_i)|$, $\zeta_2=|g_i+t{\rm sgn}(d_{i-1})|$. By taking expectation from both sides, we reach
	\begin{align}\label{eq.TVEdist2}
	&\mathds{E}\mathrm{dist}^2(\bm{g},t\partial\|\cdot\|_{\mathrm{TV}}(\bm{x}))\ge\nonumber\\
	&|\mathcal{S}_1^{+}\cup\mathcal{S}_1^{-}|+\sum_{i\in\mathcal{S}_1^{+}\cup\mathcal{S}_1^{-}}t^2({\rm sgn}(d_i)-{\rm sgn}(d_{i-1}))^2\nonumber\\
	&+\sum_{i\in\mathcal{S}_2}\mathds{E}(\zeta_1-t)_{+}^2+\sum_{i\in\mathcal{S}_2^{\prime}}\mathds{E}(\zeta_2-t)_{+}^2+\nonumber\\
	&\sum_{i\in\mathcal{S}_4}\mathds{E}(|g_i|-t-t)_{+}^2+(1+t^2)1_{1\in\mathcal{S}}\nonumber\\
	&\mathds{E}(|g_1|-t)_{+}^2+(1+t^2)1_{n-1\in\mathcal{S}}+\mathds{E}(|g_n|-t)_{+}^2.
	\end{align}
	In what follows, we calculate the expressions within \eqref{eq.TVEdist2}. First, consider $\mathds{E}(\zeta_1-t)_{+}^2$, which is calculated as follows:
	\begin{align}\label{eq.EzetalTV}
	&\mathds{E}(\zeta_1-t)_{+}^2=2\int_{0}^{\infty}a\mathds{P}(\zeta_1\ge a+t)da\nonumber\\
	&=2{\frac{1}{\sqrt{2\pi}}}\int_{0}^{\infty}\int_{t+a}^{\infty}a~(e^{-\frac{(u-t)^2}{2}}+e^{-\frac{(u+t)^2}{2}})du ~da\nonumber\\
	&=2{\frac{1}{\sqrt{2\pi}}}\int_{t}^{\infty}\int_{0}^{u-t}a~(e^{-\frac{(u-t)^2}{2}}+e^{-\frac{(u+t)^2}{2}})da~du\nonumber\\
	&=\phi_1(t,t),
	\end{align}
	where in (\ref{eq.EzetalTV}), the order of integration is changed together with a change of variable. Similarly, we have: $\mathds{E}(\zeta_2-t)_{+}^2=\phi_1(t,t)$. Also,
	\begin{align}\label{eq.Ezetal1}
	&\mathds{E}(|g_i|-t)_{+}^2=2\int_{0}^{\infty}a\mathds{P}(|g_i|\ge a+t)da\nonumber\\
	&=2\sqrt{\frac{2}{\pi}}\int_{0}^{\infty}\int_{t+a}^{\infty}a~e^{-\frac{u^2}{2}}du ~da\nonumber\\
	&=2\sqrt{\frac{2}{\pi}}\int_{t}^{\infty}\int_{0}^{u-t}a~e^{-\frac{u^2}{2}}da~du:=\phi_2(t),
	\end{align}
	where in the third line, the order of integration is changed. Thus, we have, $\mathds{E}(|g_i|-t-t)_{+}^2=\phi_2(2t)$. As a consequence, (\ref{eq.TVEdist2}) becomes:
	\begin{align}\label{eq.TVEdist2asli}
	&\mathds{E}\mathrm{dist}^2(\bm{g},t\partial\|\cdot\|_{\mathrm{TV}}(\bm{x}))\ge\nonumber\\
	&|\mathcal{S}_1^{+}\cup\mathcal{S}_1^{-}|+\sum_{i\in\mathcal{S}_1^{+}\cup\mathcal{S}_1^{-}}t^2({\rm {\rm sgn}}(d_i)-{\rm sgn}(d_{i-1}))^2\nonumber\\
	&+\sum_{i\in\mathcal{S}_2}\phi_1(t,t)+\sum_{i\in\mathcal{S}_2^\prime}\phi_1(t,t)+\nonumber\\
	&\sum_{i\in\mathcal{S}_4}\phi_2(2t)+(1+t^2)1_{1\in\mathcal{S}}+\phi_2(t)1_{1\in\overline{\mathcal{S}}}\nonumber\\
	&+(1+t^2)1_{n-1\in\mathcal{S}}+\phi_2(t)1_{n-1\in\overline{\mathcal{S}}}:=\Psi_t(s_1^{+},s_1^{-},s_2,s_3).
	\end{align}
	Finally by setting $s_3=1_{1\in\mathcal{S}}+1_{n-1\in\mathcal{S}}$, $s_2=|\mathcal{S}_2|+|\mathcal{S}_2^\prime|$, $2-s_3=1_{1\in\overline{\mathcal{S}}}+1_{n-1\in\overline{\mathcal{S}}}$, $s_1^+=|\mathcal{S}_1^{+}|$, $s_1^{-}=|\mathcal{S}_1^{-}|$, and $|\mathcal{S}_4|=n-2-s_1^{+}-s_1^{-}-s_2$, we reach
	\begin{align}
	\delta(\mathcal{D}(\|\cdot\|_{\rm TV},\bm{x}))\ge \inf_{t\ge0}\Psi_t(s_1^{+},s_1^{-},s_2,s_3):=\widehat{m}_{\rm TV}.
	\end{align}
We know from \cite[Theorems 7.1, 6.1]{amelunxen2013living} that if $$m\le \delta(\mathcal{D}(\|\cdot\|_{\rm TV},\bm{x})):=\delta,$$ then,
\begin{align}
\mathds{P}[\bm{x}~\text{is the unique solution of}~\mathsf{P}_{\rm TV}]\le 4 {\rm e}^{-\frac{(\delta-m)^2}{16\delta}}.
\end{align}
Since the function $f:z\rightarrow 4{\rm e}^{-\frac{(z-m)^2}{16z}}$ is decreasing, and the fact that $m\le \widehat{m}_{\rm TV}\le \delta$, it holds that
\begin{align}
&\mathds{P}[\bm{x}~\text{is the unique solution of}~\mathsf{P}_{\rm TV}]\le 4 {\rm e}^{-\frac{(\delta-m)^2}{16\delta}}\le\nonumber\\
& 4 {\rm e}^{-\frac{(\widehat{m}_{\rm TV}-m)^2}{16\widehat{m}_{\rm TV}}}.
\end{align}
\end{proof}
\ifCLASSOPTIONcaptionsoff
  \newpage
\fi
\bibliographystyle{ieeetr}
\bibliography{mypaperbibe}
\vskip 0pt plus -1fil
\begin{IEEEbiographynophoto}{\textbf{Sajad Daei}} 
	received the B.Sc. degree in electronic engineering from Guilan University, Guilan, Iran, in 2011, and the M.Sc. degree in communication engineering from Sharif University of Technology (SUT), Tehran, Iran, in 2013. He is currently pursuing his Ph.D. at Iran University of Science \& Technology (IUST). His main research interests include convex optimization, compressed sensing and super resolution.
\end{IEEEbiographynophoto}
\vskip 0pt plus -1fil
\begin{IEEEbiographynophoto}{\textbf{Farzan Haddadi}} was born in 1979. He received his B.Sc., M.Sc., and Ph.D. degrees in communication systems in 2001, 2003, and 2010, respectively, from Sharif University of Technology, Tehran, Iran. He joined Iran University of Science \& Technology faculty in 2011. His main research interests are array signal processing, statistical signal processing, subspace tracking, and compressed sensing.
\end{IEEEbiographynophoto}
\vskip 0pt plus -1fil
\begin{IEEEbiographynophoto}{\textbf{Arash Amini}} 
	received the B.Sc., M.Sc., and Ph.D. degrees in electrical engineering (communications and signal processing) and the B.Sc. degree in petroleum engineering (reservoir) from the Sharif University of Technology, Tehran, Iran, in 2005, 2007, 2011, and 2005, respectively. He was a Researcher with the \'Ecole Polytechnique f\'ed\'erale de Lausanne, Lausanne, Switzerland, from 2011 to 2013, working on statistical approaches toward modeling sparsity in continuous-domain. He joined Sharif University of Technology as an assistant professor in 2013, where he is now an associate professor since 2018. He has served as an associate editor of IEEE Signal Processing Letters from 2014 to 2018. His research interests include various topics in statistical signal processing, specially compressed sensing.			
\end{IEEEbiographynophoto}
\end{document}